\documentclass[twocolumn,5p,12pt]{elsarticle}
\usepackage{amssymb}
\usepackage{amsmath}
\usepackage{amsthm}
\usepackage{graphicx}
\usepackage{dcolumn}
\usepackage{paralist}
\usepackage{caption}
\usepackage{epstopdf}
\usepackage{epsfig}
\usepackage{natbib}
\usepackage{textcase}
\usepackage{bm}
\usepackage{url}
\usepackage{makecell}
\usepackage{tabularx}
\usepackage{multirow}
\usepackage{array}
\usepackage{ragged2e}
\newcolumntype{P}[1]{>{\RaggedLeft\arraybackslash}p{#1}}
\newcolumntype{C}[1]{>{\centering\let\newline\\\arraybackslash\hspace{0pt}}m{#1}}
\usepackage{wrapfig}
\usepackage{float}
\usepackage[number-unit-separator=~]{siunitx} 
\usepackage{threeparttable,booktabs}
\usepackage{xurl}

\usepackage{lineno}
\journal{Journal of Materials Research and Technology}

\makeatletter\def\Hy@Warning#1{}\makeatother 
\usepackage{microtype} 

\usepackage[]{hyperref}
\usepackage{xcolor}
\definecolor{ElsevierLink}{RGB}{33, 150, 209}
\hypersetup{
	colorlinks=true,
}

\DeclareCaptionFormat{figures}{\textbf{#1#2} {#3}}
\DeclareCaptionFormat{tables}{\textbf{#1#2} {#3}}
\begin{document}
\hypersetup{
	linkcolor = ElsevierLink,
	citecolor = ElsevierLink,
	urlcolor = ElsevierLink
}

\begin{frontmatter}

\title{Exploring diffusion bonding of niobium and its alloys with tungsten and a molybdenum alloy for high-energy particle target applications}

\author[label1]{Tina Griesemer}
 \ead{tina.griesemer@cern.ch}
\author[label1]{Rui Franqueira Ximenes}
 \ead{rui.franqueira.ximenes@cern.ch}
\author[label1]{Claudia Ahdida}
\author[label1]{Gonzalo Arnau Izquierdo}
\author[label1]{Ignacio Aviles Santillana}
\author[label2]{Jack S. Callaghan} 
\author[label1]{Gerald Dumont}
\author[label3]{Thomas Dutilleul}
\author[label1]{Adria Gallifa Terricabras}
\author[label1]{Stefan Höll}
\author[label1]{Richard Jacobsson}
\author[label3]{William Kyffin}
\author[label2]{Abdullah Al Mamun}
\author[label1]{Giuseppe Mazzola}
\author[label1]{Ana Teresa Pérez Fontenla}
\author[label1]{Oscar Sacristan De Frutos}
\author[label1]{Luigi Salvatore Esposito}
\author[label1]{Stefano Sgobba}
\author[label1]{Marco Calviani}
 \ead{marco.calviani@cern.ch}

\affiliation[label1]{European Organization for Nuclear Research (CERN), 1211 Geneva 23, Switzerland}
\affiliation[label2]{Bangor University, Bangor LL57 2DG, United Kingdom}
\affiliation[label3]{Nuclear AMRC, Rotherham S60 5WG, United Kingdom}

\begin{abstract}
Particle-producing targets in high-energy research facilities are often made from refractory metals, and they typically require dedicated cooling systems due to the challenging thermomechanical conditions they experience. However, direct contact of water with target blocks can induce erosion, corrosion, and embrittlement, especially of tungsten (W). One approach to overcoming this problem is cladding the blocks with tantalum (Ta). Unfortunately, Ta generates high decay heat when irradiated, raising safety concerns in the event of a loss-of-cooling accident.
This study explored the capacity of niobium (Nb) and its alloys to form diffusion bonds with W and TZM (a molybdenum alloy with titanium and zirconium). This is because the Beam Dump Facility (BDF), a planned new fixed-target installation in CERN's North Area, uses these target materials.
The bonding quality of pure Nb, Nb1Zr, and C103 (a Nb alloy with 10\% hafnium and 1\% titanium) with TZM and W obtained using hot isostatic pressing (HIP) was evaluated. The effects of different HIP temperatures and the introduction of a Ta interlayer were examined.
Optical microscopy indicated promising bonding interfaces, which were further characterized using tensile tests and thermal-diffusivity measurements. Their performance under high-energy beam impact was validated using thermomechanical simulations. C103 exhibited higher interface strengths and safety factors than Ta2.5W, positioning it as a potential alternative cladding material for the BDF production target.
The findings highlight the viability of Nb-based materials, particularly C103, for improving operational safety and efficiency in fixed-target physics experiments; however, considerations regarding the long half-life of \textsuperscript{94}Nb require further attention.
\end{abstract}


\end{frontmatter}

\section{Introduction}
Particle-producing targets are used in high-energy physics experiments to generate secondary particles when struck by a primary particle beam. Because they operate under harsh thermomechanical conditions, these targets often need to be made from specially tailored materials. The high power deposited in a target by a primary beam is often dissipated using a water-cooling system; however, when exposed to direct contact with water, some refractory metals experience corrosion, erosion, and embrittlement ~\cite{lillard2000corrosion}. One approach to overcoming this problem is to clad the target blocks with tantalum (Ta), and this was the method used in the KENS~\cite{kawai2001fabrication}, LANSCE~\cite{nelson2012fabrication}, ISIS~\cite{dey2018strategies}, and CSNS~\cite{wei2021advance} experiments. This cladding technique was also employed for the baseline design of the recently approved CERN Beam Dump Facility (BDF) target, which will serve the Search for Hidden Particles (SHiP) experiment~\cite{ahdida2019sps}.

The BDF target is designed to operate with a high-energy proton beam of \SI{400}{GeV/\textit{c}}, delivering \SI{4e13}{} protons per pulse and a projected accumulated \SI{4e19}{} protons on target annually. The physics of the SHIP experiment require the target materials to have high density, high atomic and mass numbers, and a short nuclear interaction length. In addition, as the target will experience an average power deposition of \SI{305}{kW}, its materials must be able to withstand challenging thermomechanical conditions. Taking these factors into consideration, TZM---a molybdenum alloy with titanium and zirconium---and pure tungsten (W) were selected as core materials for the target. Pure Ta and Ta2.5W---a Ta alloy with 2.5\% W---were considered for the cladding, and diffusion bonding between the cladding and core materials was achieved using hot isostatic pressing (HIP)~\cite{sola2019design}.

Several studies have been performed in relation to the BDF target, including characterization of its materials~\cite{FHinternal2017}, assessment of the diffusion bonding between the Ta cladding and core materials after HIP~\cite{busom2020application}, and the manufacturing, irradiation, and post-irradiation examination of a reduced-scale prototype BDF target~\cite{sola2019beam, franqueira2021jacow, griesemer2024post}. Safety concerns were raised due to the high decay heat of Ta, and this led to a study considering loss-of-cooling accidents~\cite{mena2022loca} and the exploration of alternative cladding materials. The refractory metal niobium (Nb) exhibits excellent thermomechanical properties, high ductility, reduced activation under the operational conditions of the BDF, and lower decay heat than Ta. Moreover, Nb has been found to exhibit effective bonding with both TZM and W, similar diffusivity rates to Ta, and full solubility with the target materials~\cite{basuki2011investigation, franqueira2023}.
There are currently no Nb-clad target blocks employed in research facilities; hence, the present study focused on assessing the quality of diffusion bonding between Nb-based materials and the core materials, ultimately determining their suitability as cladding for BDF target blocks.

Thermal and mechanical tests have previously been used to evaluate the bonding quality of Ta and Ta2.5W with TZM and W via HIP under different manufacturing parameters~\cite{busom2020application}. Using these previous results for Ta as a baseline, in the present work, prototypes with the same geometry were used to evaluate three different materials: (1)~pure Nb; (2)~Nb1Zr (a Nb alloy containing 1\% zirconium that has better mechanical properties than pure Nb); and (3)~C103 (a Nb alloy containing 10\% hafnium and 1\% titanium that presents superior mechanical properties). In addition, the effects on the bonding interface of using different HIP temperatures (\SI{1200}{} and \SI{1400}{\degreeCelsius}) and the use of a Ta interlayer with a thickness of \SI{50}{\micro\meter} were evaluated. Initial observations by optical microscopy (OM) showed that the prototypes had good bonding interfaces~\cite{griesemer2023jacow}. This study then sought to determine the strengths of the interfaces using tensile tests and to measure the thermal contact resistances ($R$-values) between the core and cladding materials.

The mechanical and thermal properties of the pure Nb and Nb alloys noted above were established in a previous study~\cite{Nbinternal2024}. As such, the previously conducted thermomechanical simulations of high-energy beam impact were rerun in ANSYS\textsuperscript{\textregistered} Mechanical\texttrademark~\cite{ansys}. The Nb materials were compared with each other based on their manufacturing parameters, diffusion-bonding quality, interface strength, thermal contact resistance, and the presence or absence of a Ta interlayer. Additionally, the results were compared with those from Ta-clad blocks in terms of their bonding quality, calculated safety factors under operational conditions, physics (decay heat and experimental benefits), ease of dismantling, and waste disposal.

\begin{figure}[bt]
\centering
\includegraphics[width=\columnwidth]{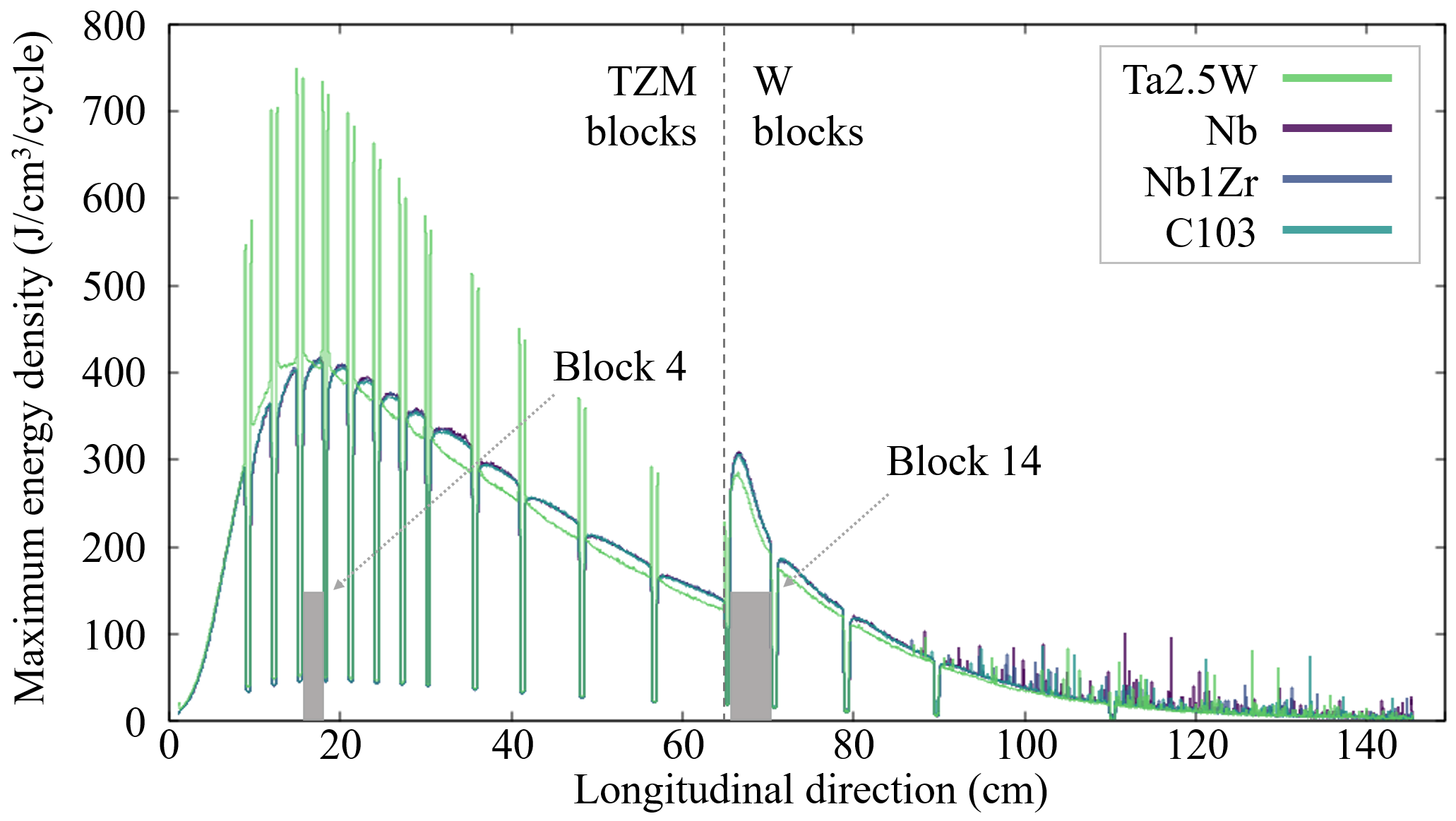}
\caption{Maximum energy density deposited along the longitudinal direction of the BDF target when using different cladding materials.}
\label{fig:energy_density}
\end{figure}

\section{Thermomechanical calculations}

\subsection{Finite-element analysis models}
\label{sim:methods}
Calculations were performed to determine whether Nb and its alloys are suitable cladding materials for the BDF target. To this end, the most critical blocks in terms of temperatures and thermally induced stresses for each core material of the baseline target design were assessed: Block~4 (TZM) and Block~14 (W). These blocks have a diameter of \SI{250}{mm}, thicknesses of \SI{25}{mm} and \SI{50}{mm}, respectively, and a cladding thickness of \SI{1.5}{mm}. The cladding materials Nb, Nb1Zr, and C103 were assessed and compared with Ta2.5W. Fig.~\ref{fig:energy_density} shows plots of the maximum deposited energy density per cycle in the longitudinal direction for BDF targets clad with Nb, Nb1Zr, C103, and Ta2.5W. It can be seen that due to their similar densities and chemical compositions, the deposited energy densities for the Nb materials were fairly similar. These values were calculated using FLUKA Monte Carlo simulations~\cite{ahdida2022new}, and they were then imported into ANSYS Mechanical~\cite{ansys} to conduct thermomechanical finite-element analysis (FEA).

First, steady-state conditions were computed to provide the initial inputs for subsequent transient thermal simulations of three successive beam impacts. Thermally induced stresses were calculated based on the temperature distributions during the third pulse in the structural simulations. The FEA models each used the same thermal and structural boundary conditions, including constant forced convection, weak-spring spatial constraints, and an absence of pre-stress conditions, such as residual stresses from the HIP process. The material properties were taken from internal material characterizations~\cite{FHinternal2017, cerninternal2022}; our previous FEA models used literature values for the Nb materials~\cite{griesemer2023jacow}. In general, the target design was based on the operation in the elastic-material range because plastic deformation of the blocks can interfere with the cladding--core bonding.

\subsection{Finite-element analysis results}
\begin{figure}[bt]
\centering
\includegraphics[width=\columnwidth]{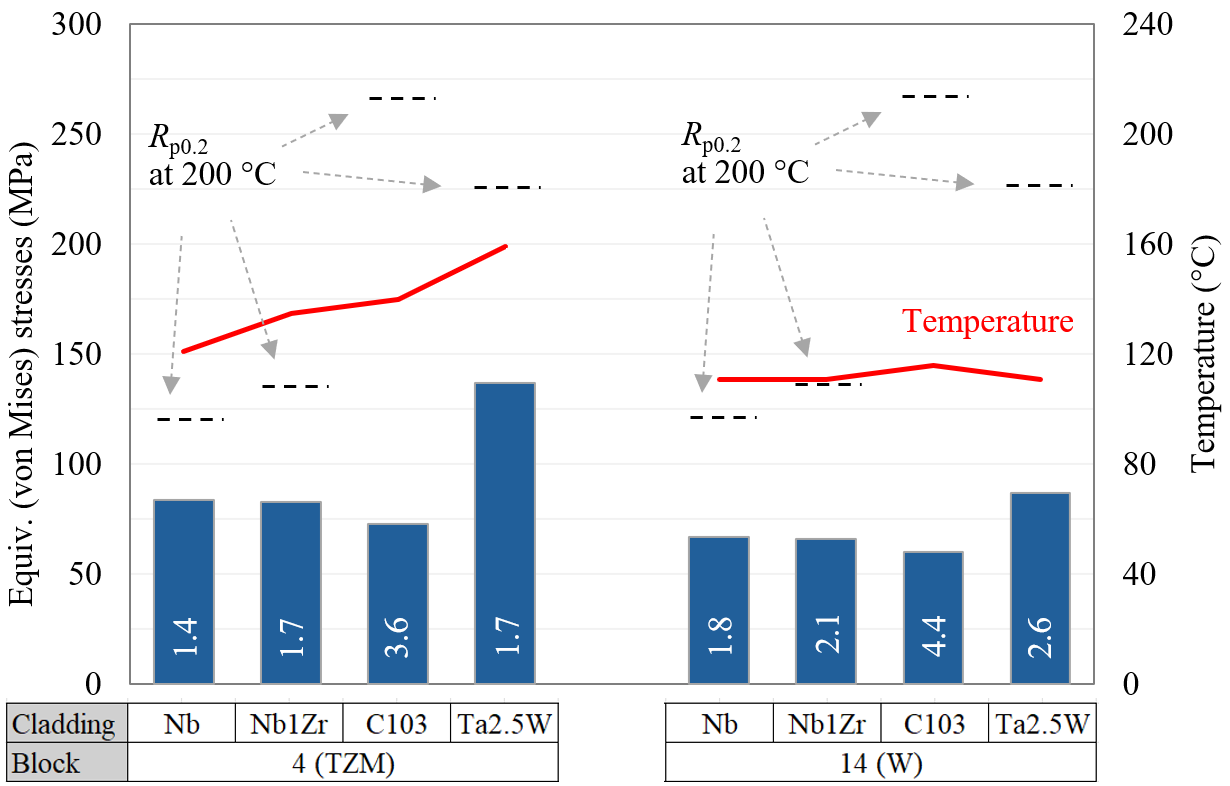}
\caption{Maximum temperatures ( in red) and equivalent (von Mises) stresses (in blue) for different cladding materials were calculated using FEA models of the BDF target. The safety factors (in white) were determined from the calculated stresses and their previously measured yield strength ($R_{\mathrm{p}0.2}$)~\cite{FHinternal2017,Nbinternal2024}. }
\label{fig:simulation}
\end{figure}

As the densities of the Nb materials are lower than that of Ta2.5W, the calculated energy deposition in all Nb-clad Block~4 (TZM) was also lower, as shown in Fig.~\ref{fig:energy_density}. In general, less energy is deposited upstream of Block~14 (W), resulting in slightly higher energy deposition in the W core when the Nb materials are used as cladding. Fig.~\ref{fig:simulation} also shows the maximum temperatures, equivalent (von Mises) stresses, and safety factors for all the cladding materials. The stresses in the Nb materials were fairly similar but lower than those in the Ta2.5W cladding; however, considering their differences in yield strength ($R_{\mathrm{p}0.2}$), Nb had the lowest safety factor and C103 had the highest for both blocks. In addition, the cladding temperatures were higher for Ta2.5W than for the Nb materials. The equivalent stress for the TZM core material was approximately \SI{120}{MPa} for all cladding materials. The W in Block~14 had maximum principal stresses of \SI{90}{MPa} with Ta2.5W cladding and approximately \SI{110}{MPa} with the Nb materials, while the cladding temperatures were fairly similar for all four cladding materials. C103 presented the highest safety factor, followed in decreasing order by Ta2.5W, Nb1Zr, and Nb. In summary, based on only temperatures and thermally induced stresses, of the cladding materials studied, C103 was found to be the best choice for the BDF target.

\section{Prototypes}
\subsection{Design and manufacturing methods}
The diffusion-bonding capabilities of TZM and W with Ta alloys were examined in a previous study by analyzing small prototype capsules bonded via HIP~\cite{busom2020application}. To allow direct comparisons between these previous results and the Nb materials considered in the present work, the same prototype geometry, diffusion-bonding procedure, and testing parameters were selected. Cylindrical prototypes were made from a TZM or W core and three cladding parts. Each prototype had a diameter of \SI{26}{mm}, a height of \SI{50}{mm}, and a cladding thickness of \SI{1}{mm} radially and \SI{10}{mm} on the flat surfaces. Because it has previously been shown that a \SI{50}{}-\unit{\micro\meter}-thick Ta interlayer can improve the quality of diffusion bonding~\cite{busom2020application}, Ta foil interlayers were included on one of the faces of each prototype. 

Two prototypes were examined for each material combination, resulting in a total of 12~prototypes for the two core materials (TZM and W) and three Nb materials (pure Nb, Nb1Zr, and C103); two Ta prototypes with a W core were also assessed as a reference to the previous study. Because material-quality issues were detected in the pure Nb (as will be discussed later), four additional Nb-clad prototypes were added to the study. Thus, a total of 18 physical prototypes were examined.

The following steps were performed to achieve diffusion bonding between the cladding and core materials. (1)~First, prototype tubes and discs were machined with a tolerance gap of \SI{0.05}{} to \SI{0.20}{mm}, producing an as-machined surface finish. (2)~This was followed by surface cleaning with isopropanol. (3)~The tubes and discs of the cladding components were then joined by electron-beam welding (EBW). After one side had been welded, the workpiece was kept under vacuum overnight, and the process was repeated on the other side. Due to the small size of the prototypes, beam deflection was used for the EBW rather than mechanical movement. This was achieved using a Pro-Beam~K25~\cite{EBW} with a vacuum below \SI{6e-4}{mbar}, a voltage of \SI{80}{kV}, a current of \SI{40}{mA}, and a welding speed of \SI{40}{mm/s}. (4)~To assess the quality of each weld, helium leak testing was performed following the BS~EN~13185 standard~\cite{BSEN13185}. The samples were each impregnated in a pressure vessel for \SI{20}{h} at \SI{4}{bar}, allowed to rest for \SI{20}{min} after being removed from the vessel, and then tested individually in a vacuum chamber using an ASM~340 leak detector (Pfeiffer Vacuum GmbH), which has a leakage sensitivity of \SI{1e-8}{mbar{\cdot}l/s}. (5)~The prototypes were separated into two groups---odd and even numbered---and then sequentially placed into a HIP furnace. Two different HIP cycles were used; both had a pressure of \SI{200}{MPa} and a dwell time of \SI{3}{h}, but a temperature of \SI{1200}{\degreeCelsius} was used for the odd-numbered prototypes (cycle~L) while \SI{1400}{\degreeCelsius} was used for the even-numbered prototypes (cycle~H). The heating rate of cycle~H was also higher (L: \SI{6.3}{K/min}; H: \SI{7.6}{K/min}), but their cooling rates were fairly similar (L: \SI{-8.7}{K/min}; H: \SI{-8.5}{K/min}). (6)~Finally, ultrasonic testing (UT) by immersion was carried out to evaluate the diffusion bonding between the cladding and core materials. Non-HIPed and non-welded prototypes were used as reference bodies for each material combination.

\subsection{Specimen extraction and testing procedure}
Each prototype was cut longitudinally, and its bonding interfaces were inspected using OM. After this, one thermal diffusivity specimen and four tensile specimens were extracted from each flat bonding interface of each prototype.
The thermal specimens each had a diameter of \SI{12.6}{mm} and a thickness of 2.3--2.35~\unit{mm}; the bonding interfaces were located close to their centers, and half of these included a \SI{50}{}-\unit{\micro \meter}-thick layer of Ta~foil. The thicknesses of the layers in each sample were calculated based on the total sample thickness, the previously derived bulk material densities, and the mixed sample density, which was measured in ethanol using a Sartorius Quintix 224-1x scale. A thin graphite coating was applied before performing laser flash analysis using a NETZSCH LFA~457 MicroFlash using a mercury-cadmium-telluride infrared detector, a laser voltage of \SI{2978}{V}, a helium atmosphere, and a temperature range from \SI{30}{} to \SI{450}{\degreeCelsius}. The thermal contact resistance ($R$-value) of each specimen was determined, and the thermal contact conductance (TCC), which is the reciprocal of the $R$-value, could then be obtained. A TCC sensitivity study was then performed using thermomechanical calculations based on the same FEA models of the Ta2.5W-clad blocks~4 and 14 as described in Section~\ref{sim:methods}. The TCC threshold for each block was established based on a \SI{10}{\degreeCelsius} temperature rise compared to the ideal bonding conditions, which would correspond to an $R$-value of zero.

The tensile specimens each had a length of \SI{16}{mm}, a thickness of \SI{1.5}{mm}, and a gauge width of \SI{2}{mm}, with the bonding interface positioned in the middle of each specimen. Due to their small size, lines were scribed on each specimen to define a gauge length of \SI{3.6}{mm} with an accuracy of 1\%. A Walter~+ Bai~AG LFMZ-100-M testing machine with a virtual extensometer based on the method of digital image correlation was used at a strain rate of \SI{2.9e-4}{s^{-1}} at room temperature (10--35~\unit{\degreeCelsius}) to derive the engineering stress based on the ISO~6892 standard.

\begin{table}[bt]
   \centering
   \caption{Overview of the materials used in all fabricated prototypes along with their leak-test results, applied HIP cycles, UT results, and bonding-quality results.}
   \renewcommand{\arraystretch}{1.2}
   \resizebox{\columnwidth}{!}{
   \begin{tabular}{p{0.75cm}p{1.9cm}p{1.8cm}p{1.2cm}p{1.2cm}p{1.6cm}p{1.7cm}}
       \toprule
       \toprule
       \textbf{No.} & \textbf{Cladding \newline material}   & \textbf{Core \newline material}  & \textbf{Leak \newline test}   &   \textbf{HIP \newline cycle}  &   \textbf{UT \newline result} & \textbf{Bonding \newline quality} \\
       \midrule
          1 & Nb          & W       & \underline{Failed}   &  L & \underline{Failed}  & \underline{Failed} \\
          2 & Nb          & W       & \underline{Failed}   &  H & Passed  & Passed \\
          3 & Nb          & TZM     & Passed   &  L & Passed  & Passed \\
          4 & Nb          & TZM     & \underline{Failed}   &  H & Passed  & Passed \\
          5 & Nb1Zr       & W       & Passed   &  L & Passed  & Passed \\
          6 & Nb1Zr       & W       & Passed   &  H & Passed  & Passed \\
          7 & Nb1Zr       & TZM     & Passed   &  L & Passed  & Passed \\
          8 & Nb1Zr       & TZM     & Passed   &  H & Passed  & Passed \\
          9 & C103        & W       & Passed   &  L & Passed  & Passed \\
          10 & C103       & W       & Passed   &  H & Passed  & Passed \\
          11 & C103       & TZM     & Passed   &  L & Passed  & Passed \\
          12 & C103       & TZM     & Passed   &  H & Passed  & Passed \\
          13 & Ta         & W       & Passed   &  L & Passed  & Passed \\
          14 & Ta         & W       & Passed   &  H & Passed  & Passed \\
          15 & Nb         & W       & Passed   &  L & Passed  & Passed \\
          16 & Nb         & W       & Passed   &  H & Passed  & Passed \\
          17 & Nb         & TZM     & Passed   &  L & Passed  & Passed \\
          18 & Nb         & TZM     & Passed   &  H & Passed  & Passed \\
       \bottomrule
       \bottomrule
   \end{tabular}
   }
   \label{tab:bonding_results}
\end{table}

\subsection{Diffusion-bonding quality}
\begin{figure}[bt]
\centering
\includegraphics[width=\columnwidth]{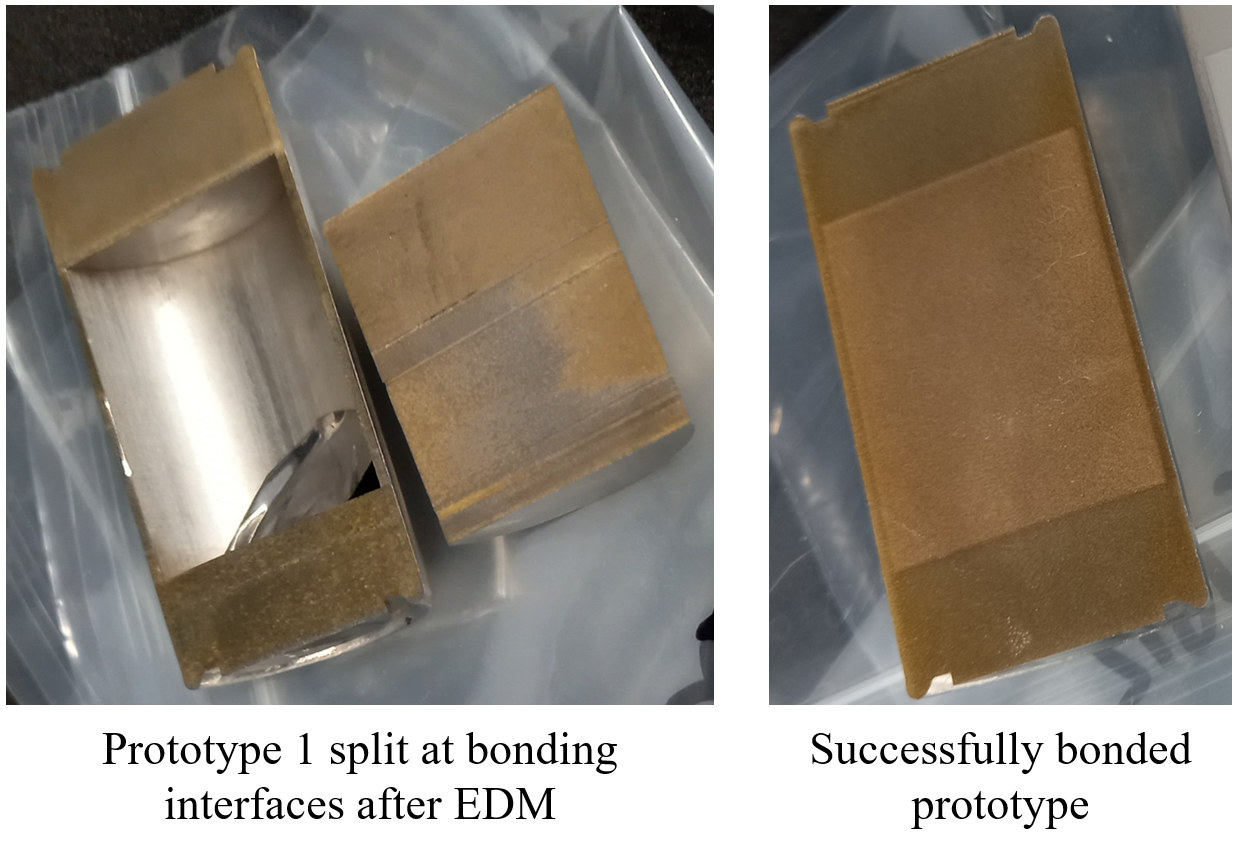}
\caption{Prototype~1 showed no signs of bonding and fell apart after the first EDM cut; it also failed the UT.}
\label{fig:failed_capsule}
\end{figure}

Table~\ref{tab:bonding_results} lists the bonding-quality results for each prototype. The Nb-clad prototypes $1$, $2$, and $4$ failed the leak test, and as noted above, a material defect was detected using dye penetrant testing (PT). The observed defects in prototypes~$1$--$4$ were patched by local welding, and four new Nb-clad prototypes, prototypes~$15$--$18$ were added to the study. The helium leak tests were not repeated for prototypes~$1$--$4$. While all other prototypes passed UT, prototype~$1$ showed discontinuity at its interfaces. When all prototypes were subsequently cut longitudinally, prototype~$1$ immediately fell apart at the interfaces and did not show any signs of bonding between the core and cladding (see Fig.~\ref{fig:failed_capsule}, left). The other prototypes were inspected by OM, and no others showed detachment at the bonding interfaces. The Nb materials thus demonstrated enough ductility to deform using the chosen HIPing parameters.

As the helium leak testing of prototypes~$1$--$4$ was not repeated after patching, it is possible that the weld patching was not successful in prototype~$1$, meaning that the cladding may not have been airtight. In these circumstances, the area between the cladding and core would not have been in vacuum but at the same isostatic pressure as the HIPing furnace, resulting in the interfaces not being brought together to be joined.

\subsection{Results of thermal testing}
\begin{figure*}[h!]
\centering
\includegraphics[width=1.65\columnwidth]{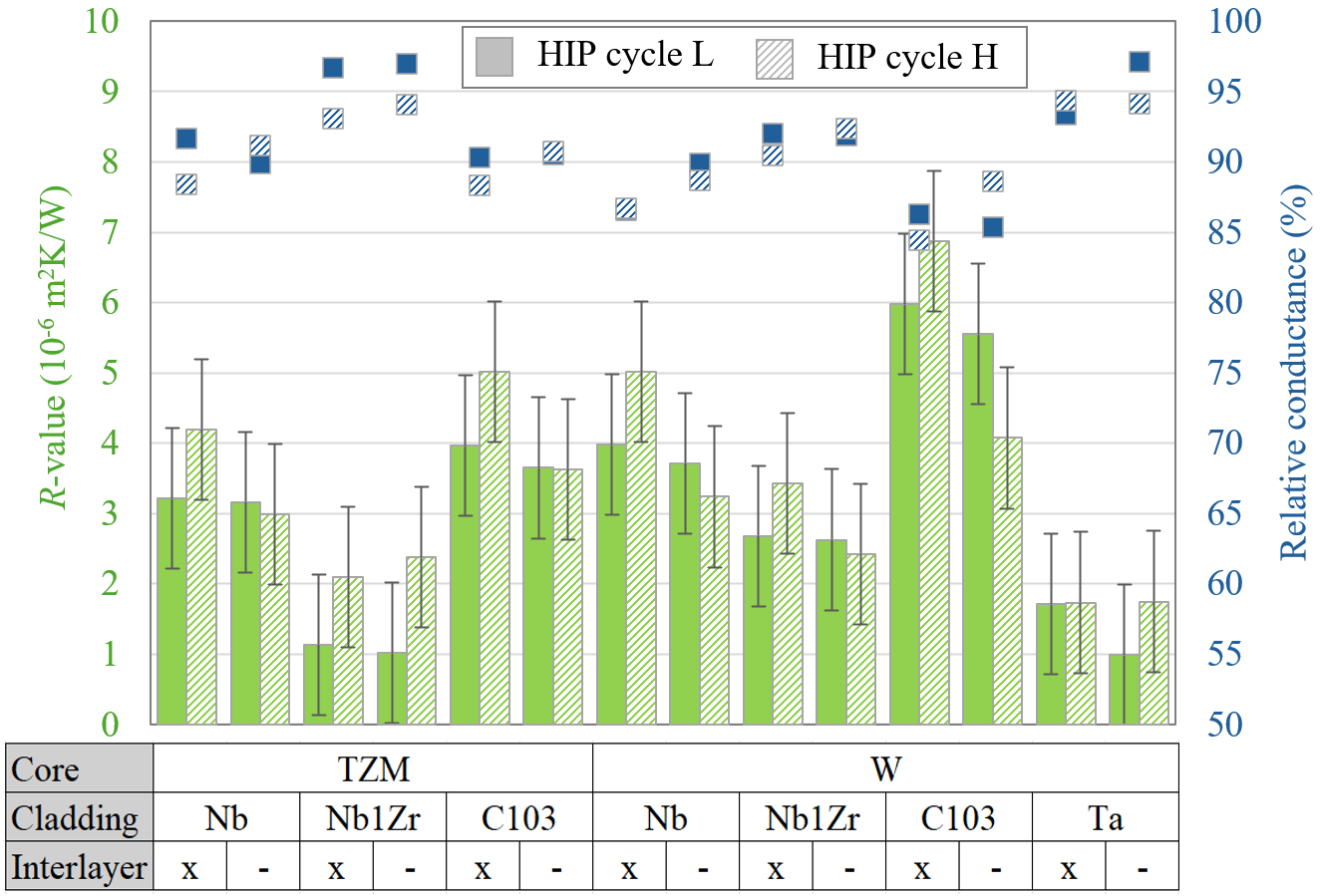}
\caption{Thermal contact resistance ($R$-value) for each material combination and its relative conductance based on its theoretically calculated conductance at \SI{30}{\degreeCelsius}.}
\label{fig:thermal_comp}
\end{figure*}

The $R$-values and relative conductivities of each thermal specimen at room temperature are presented in Fig.~\ref{fig:thermal_comp}. The relative conductivities were calculated using the measured conductance and the theoretical conductance based on the individual geometrical measurements and layer thicknesses of each sample. The $R$-values were in the range \SI{1e-6}{} to \SI{6e-6}{m^{2}\,K\,W^{-1}} and the relative conductance of every specimen was greater than 85\%. No clear dependencies on the presence/absence of Ta foil or the HIP cycle were detected. Nb1Zr appeared to have the best thermal contact, while C103 had the worst, and slightly lower $R$-values were observed for TZM than for W; however, these observations were not very pronounced due to high level of dispersion in the computed $R$-values. In addition, when compared with a theoretically perfectly bonded interface, the TCC study of the Ta2.5W-clad blocks revealed a temperature increase of \SI{10}{\degreeCelsius} for bonding interfaces with a TCC of \SI{4e4}{W\,m^{-2}\,K^{-1}}, implying an $R$-value of \SI{2.5e-5}{m^{2}\,K\,W^{-1}} for both blocks. All measured $R$-values were less than a quarter of the determined threshold for a temperature increase of \SI{10}{\degreeCelsius}.


\subsection{Tensile testing results}
\begin{figure}[bt]
\centering
\includegraphics[width=\columnwidth]{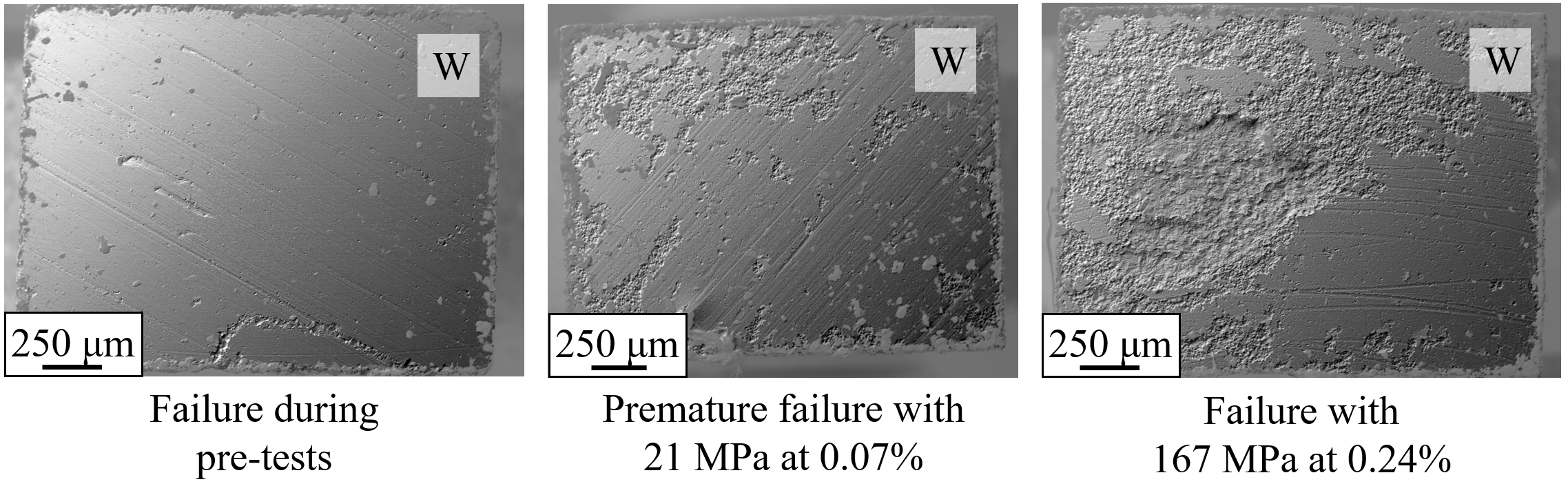}
\caption{Scanning electron microscopy images taken from the fracture interfaces of Ta/W tensile specimens that fractured at difference strengths.}
\label{fig:interface_frcatures}
\end{figure}

During extraction, some tensile specimens split immediately at the bonding interface, resulting in 104 of the 136 specimens being successfully extracted. To calibrate the machine, the remaining eight Ta/W specimens were used from the intended  16. Five samples fractured during their installation in the machine before testing. One specimen showed premature failure, and the remaining two fractured at \SI{167}{MPa} (cycle~L) and at \SI{185}{MPa} (cycle~L with Ta foil) with the interface area of $1.5 \times 2.0$~mm$^2$. In the previous study, the Ta/W specimens fractured at \SI{190}{MPa} (\SI{1200}{\degreeCelsius}, \SI{150}{Pa}) and at \SI{155}{MPa} (\SI{1400}{\degreeCelsius}, \SI{200}{Pa})~\cite{busom2020application}. Fig.~\ref{fig:interface_frcatures} shows scanning electron microscopy images of fracture interfaces occurring at different bonding strengths. While the interface of the specimen that failed during pre-tests shows no signs of bonding, the interfaces with higher bonding strengths seemed to have fractured inside the W bulk material, indicating that these interfaces had greater strength than the W bulk material.

\begin{figure*}[h!]
\centering
\includegraphics[width=1.45\columnwidth]{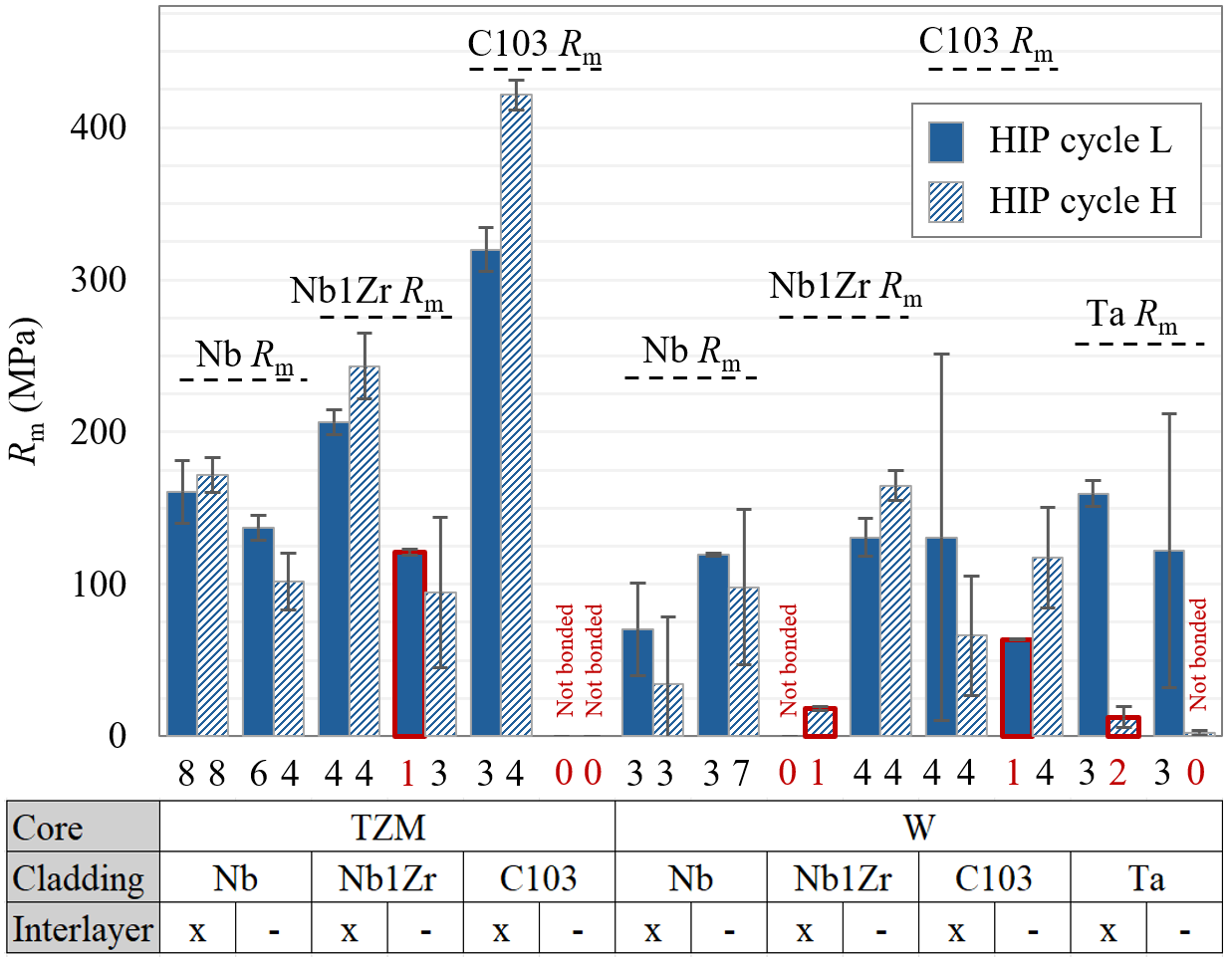}
\caption{Interface strengths of the tensile specimens with the number of available specimens per parameter combination and the interface area of $1.5 \times 2.0$~mm$^2$ compared to the $R_{\mathrm{m}}$ values of the cladding materials~\cite{FHinternal2017, Nbinternal2024}. The error bars represent the standard deviations.}
\label{fig:tensile}
\end{figure*}

The breaking strengths of all of the tensile specimens are shown in Fig.~\ref{fig:tensile}. Many of the specimens broke during extraction or during installation in the tensile testing machine, or they underwent premature failure. Hence, there were generally only a few specimens available for each parameter combination; this resulted in large error bars. In general, the TZM specimens presented higher interface strengths than the W specimens. Their interface strengths seemed to increase and were correlated with the higher tensile strengths $R_{\mathrm{m}}$ of the cladding materials. Considering that the $R_{\mathrm{m}}$ value of TZM is \SI{618}{MPa} at 22\,\unit{\degreeCelsius}~\cite{FHinternal2017}, the Nb materials were the weaker bulk materials. In addition, the TZM specimens showed higher interface strengths for all Nb materials when a Ta interlayer was used; indeed, the bonding with C103 was unsuccessful in the absence of Ta foil. Nonetheless, the C103/TZM samples that included a Ta interlayer and underwent cycle~H had the highest interface strength of all. Higher stresses were achieved using HIP cycle~H when Ta foil was included with the TZM specimens, whereas higher values were obtained without Ta foil when using cycle~L.

The W tensile specimens did not show the same behavior as the TZM samples, and the interface strength did not correlate with the $R_{\mathrm{m}}$ value of the cladding material. The interface strength was higher for Nb and Nb1Zr when no foil was used. HIP cycle~L resulted in higher values for Nb, while cycle~H led to higher values for Nb1Zr. The C103/W samples did not display a clear tendency regarding which HIP cycle resulted in greater strength or whether the strength was affected by the use of Ta foil; however, a C103/W sample with a Ta interlayer and cycle~L achieved the highest interface strength of all the W samples. Nonetheless, the highest mean interface strength was measured from Nb1Zr/W samples bonded using cycle~H. The Ta/W specimens did not bond when using HIP cycle~H, and when using cycle~L, the interface strength was greater when Ta foil was included.

It is notable that the W specimens presented lower interface strengths than their TZM counterparts. The tensile strength $R_{\mathrm{m}}$ of W is \SI{387}{MPa} for forged, recrystallized W at 20\,\unit{\degreeCelsius}~\cite{mpdb}, lower than the values for TZM and C103 (\SI{438}{MPa}) at the same temperature~\cite{Nbinternal2024}. In addition, nonuniform diffusion bonding between the cladding and core materials was observed. If the small bonding interfaces ($1.5 \times 2.0$~mm) were extracted in areas that did not bond well, little to no interface strength was observed; however, the thermal specimens, which each had a diameter of \SI{12.6}{mm}, did not present the same behavior.

\section{Material-quality issue in Nb bulk material}
\subsection{Investigation steps}
\begin{figure}[bt]
\centering
\includegraphics[width=\columnwidth]{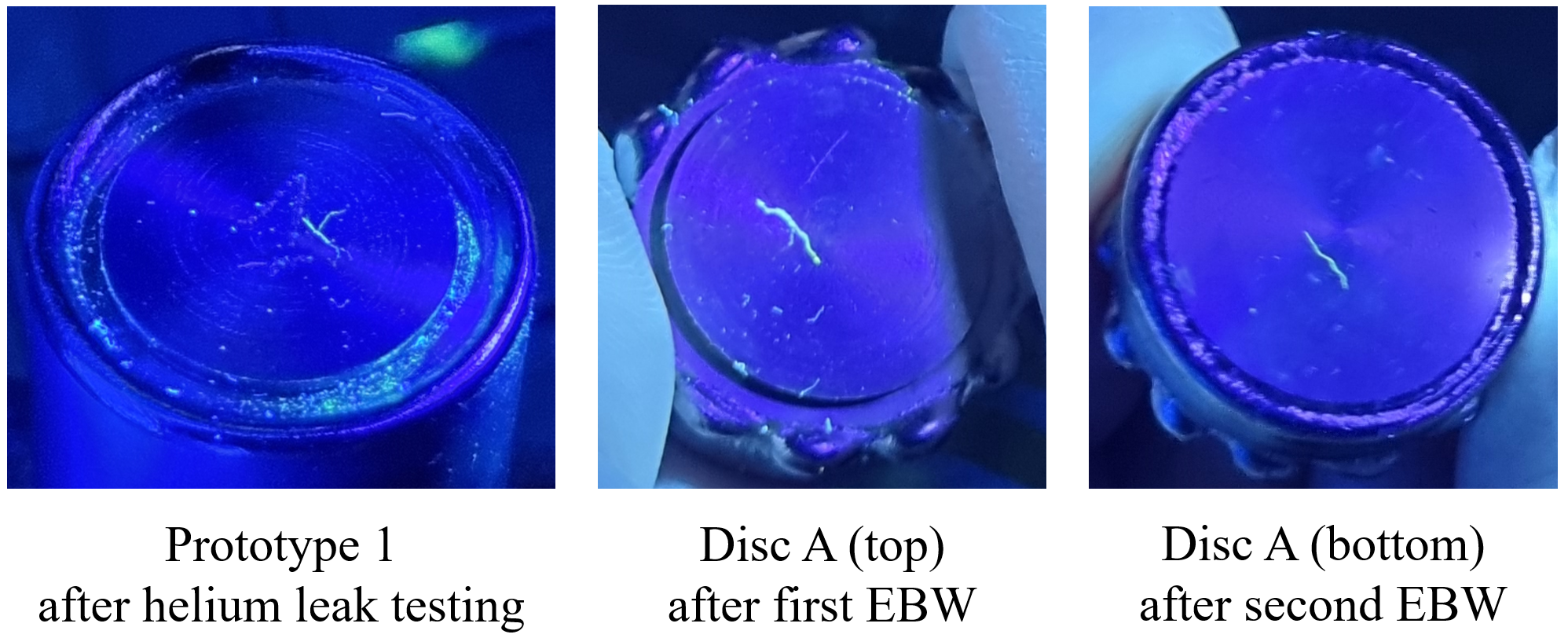}
\caption{Cracks revealed in pure Nb discs using dye penetrant testing.}
\label{fig:PT_cracks}
\end{figure}

As noted above, during the helium leak testing, three Nb-clad prototypes failed. Dye PT revealed cracks in the centers of most of the Nb discs (see Fig.~\ref{fig:PT_cracks}). Even prototype~$3$, which passed the leak test, exhibited a crack in one disc. Two pure Nb discs---Disc~A, which was from the same material batch as prototypes~$1$--$4$, and Disc~B, which was from the new batch of prototypes~$15$--$18$---with a diameter of \SI{24}{mm} and a thickness of \SI{10}{mm} were used for further examination. Both were tested by dye PT, and EBW was conducted on the outer diameter (bottom). Dye PT was then redone, EBW was repeated on the other side (top), and final dye PT was performed.

\begin{figure}[bt]
\centering
\includegraphics[width=\columnwidth]{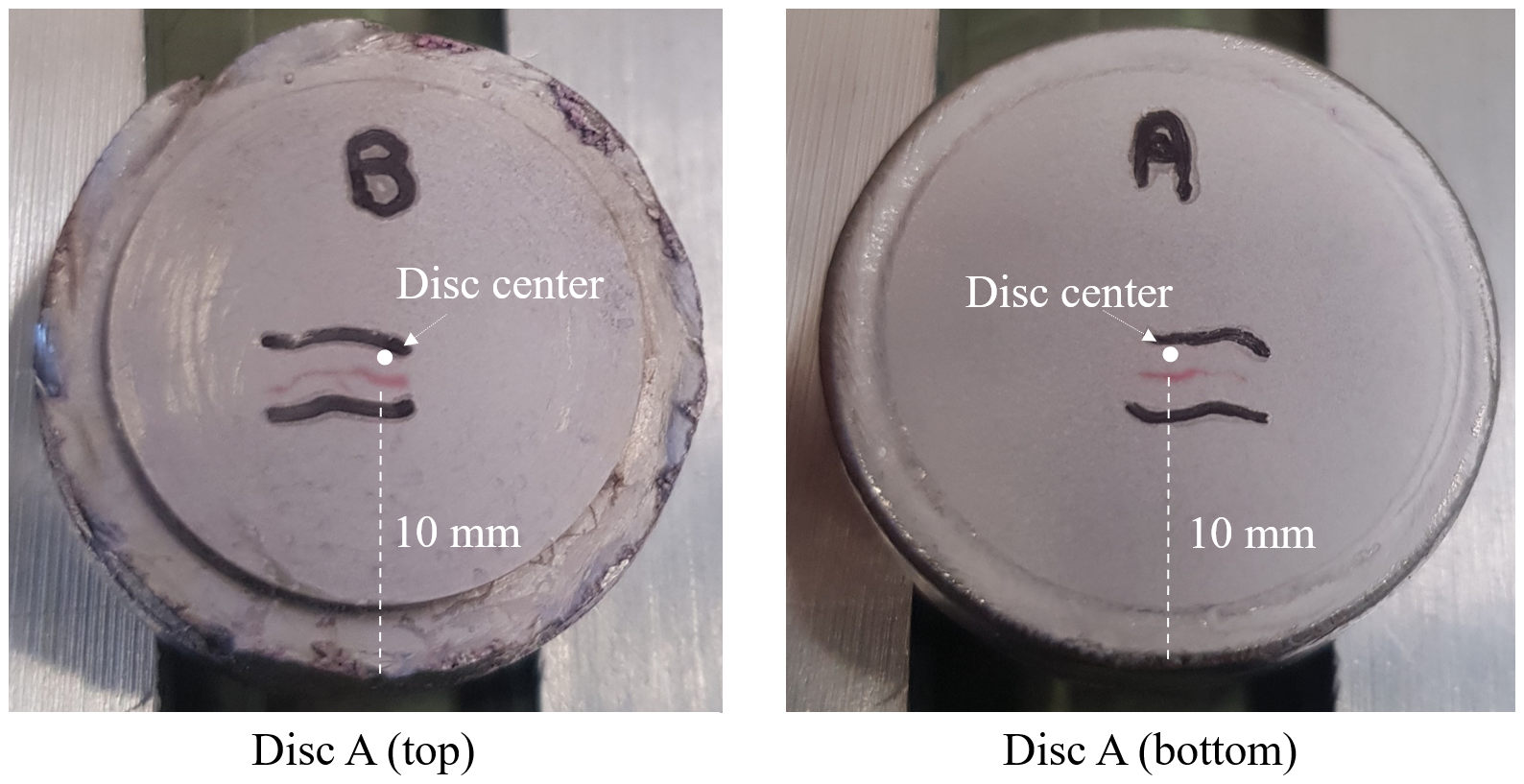}
\caption{The cracks on both sides of Disc~A (in red) are approximately \SI{5}{mm} long and were indicated with black lines on either side after dye penetrant testing. Disc~A has the same orientation in both images.}
\label{fig:marked_disc}
\end{figure}

\begin{figure*}[bt]
\centering
\includegraphics[width=1.8\columnwidth]{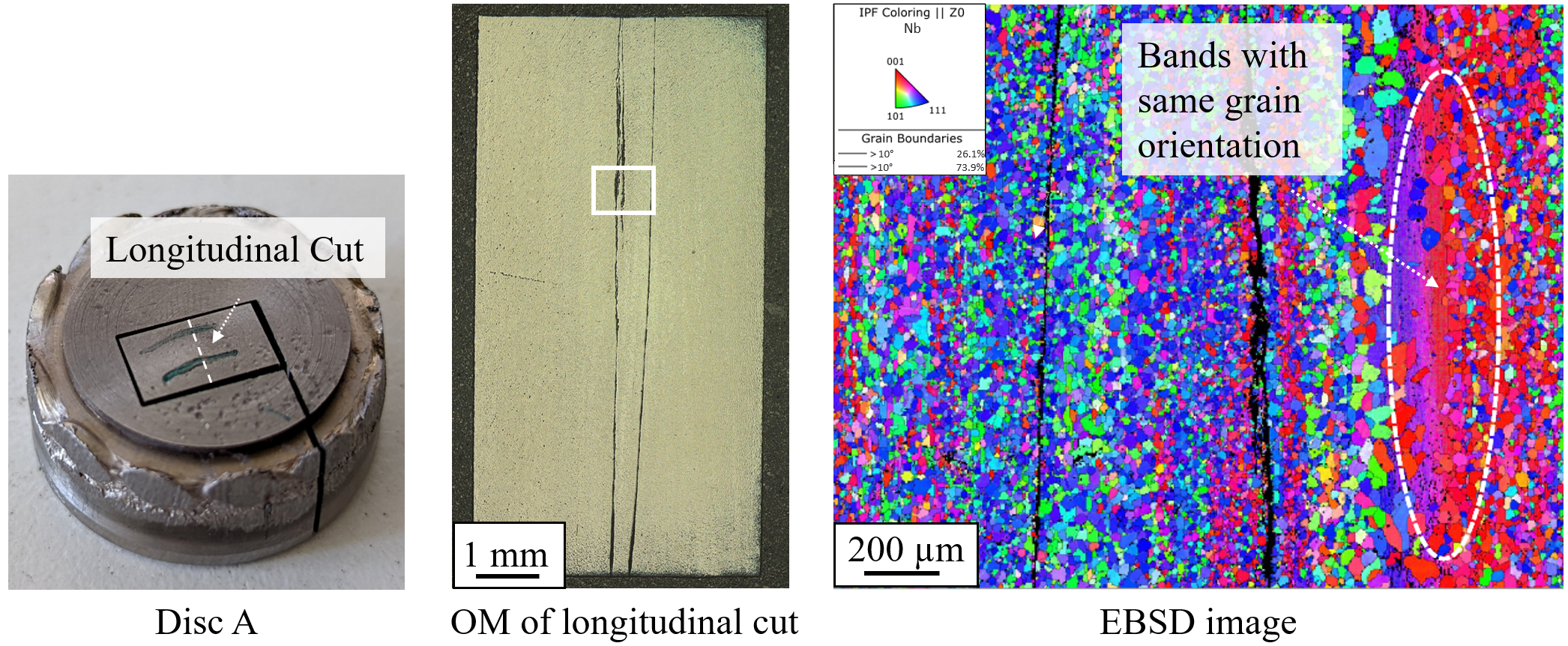}
\caption{Photograph, optical microscopy (OM) image, and EBSD IPF map of cracks detected in Disc~A.}
\label{fig:crack_EBSD}
\end{figure*}

Although Disc~B showed no PT indications, Disc~A exhibited a crack on the side opposite the welding (top), and after the second EBW, another crack appeared, again on the opposite side to the weld (bottom), as shown in Fig.~\ref{fig:PT_cracks}. Considering the high ductility of Nb~\cite{zamiri2006mechanical} and the distance of approximately \SI{10}{mm} to the weld, cracking due solely to the thermal effects of EBW was excluded. Computed tomography and UT were conducted to detect existing cracks in Disc~A; however, both methods were unsuccessful. Dye PT was redone, and the positions of the cracks were highlighted on both sides (see Fig.~\ref{fig:marked_disc}). Both cracks had a similar size, position, and orientation, and they were located on top of each other. A small piece containing these cracks was extracted from Disc~A, and this was cut perpendicular to the cracks. The piece was inspected by OM and electron backscatter diffraction (EBSD) along the longitudinal cut (see Fig.~\ref{fig:crack_EBSD}). The OM image shows multiple cracks running through the whole thickness. In addition, the EBSD inverse pole figure (IPF) maps show bands of grains with the same crystallographic orientation and not fully recrystallized.

\subsection{Origin of quality issue in Nb}
In summary, the cracking behavior was only observed in Nb discs from the same manufacturing batch, and these cracks went through the whole thickness along the disc axis. The supplier stated that these discs were taken from a rod, and they were turned afterwards. All of the cracks detected in the Nb discs showed a similar shape and location. Rod-quality issues are known to occur along the axis in the center, and they are commonly caused by insufficient temperature and pressure during the manufacturing - e.g. forging - process~\cite{sgobba2023failure}, leading to non-recrystallized zones with poor mechanical properties. Furthermore, it was assumed that the turning of the discs after cutting will have smeared material into the cracks and covered them at the surface level. Hence, the cracks in Disc~A were not detected by PT, but the thermally induced stresses that occurred during EBW caused them to be revealed in subsequent PT.

In addition, the EBSD IPF images showed bands of grains with the same orientation and non-recrystallized zones, parallel to the detected cracks (see Fig.~\ref{fig:crack_EBSD}). It is suspected that these bands assisted the formation of the observed cracks. Similar layered microstructure - i.e. bands with different crystallographic orientation - is often observed in Nb bulk materials like sheets and plates ~\cite{gallifa2023jacow}.

\section{Discussion}
\subsection{Performance of Nb materials}
The three Nb materials were fairly similar in terms of bonding quality observed by OM, thermal interface resistance ($R$-value), physical advantages for the target experiment, and radiation protection considerations, including residual dose rates, decay heat, Swiss clearance limits (LL)~\cite{SwissRPO}, and IAEA transport limits (A2)~\cite{IAEAA2}. The survival rate of each material combination was defined as the ratio of the number of tensile specimens tested to the number extracted from the prototypes. The pure-Nb tensile specimens had the highest survival rates, at 81\% with TZM and 50\% with W, while Nb1Zr and C103 had a range of 20\%--40\% for TZM and W. C103 presented the highest interface strength, safety factor, and $R_{\mathrm{m}}$, followed by Nb1Zr and Nb in turn. This means that of the three Nb materials tested, C103 is the best candidate cladding material for the BDF target.

The use of a Ta interlayer did not result in any notable differences during thermal testing. The interface strengths and survival rates of all three Nb materials bonded with TZM were higher when using Ta foil, and C103 did not bond with TZM at all without using the foil. When bonded with W, pure Nb and Nb1Zr presented higher interface strengths and survival rates without Ta foil; the C103/W specimen did not exhibit any clear tendency in this regard.

\subsection{Cladding candidate for BDF target}
A detailed comparison of Ta2.5W and C103 was performed to determine which would be a more suitable cladding material for the BDF target. Both alloys achieved good thermal bonding between the cladding and core materials, and the resulting thermal contact resistance ($R$-value) was concluded to be insignificant. The FEA models showed similar cladding temperatures for Block~14, and they were slightly lower for C103-clad Block~4. The calculated safety factors were higher for C103, at 3.6 and 4.4 for blocks~4 (TZM) and 14 (W), respectively, compared to corresponding values of 1.7 and 2.6 for Ta2.5W cladding. In addition, the tensile strength $R_{\mathrm{m}}$ at \SI{200}{\degreeCelsius} is \SI{366}{MPa} for C103 and \SI{331}{MPa} for Ta2.5W~\cite{FHinternal2017,cerninternal2022}. The interface strength was significantly higher for C103/TZM (\SI{420}{MPa}) than for Ta2.5W/TZM (\SI{325}{MPa})~\cite{busom2020application}. Neutron-irradiated Ta alloys generate a higher decay heat than Nb (and its alloys) due to the production of \textsuperscript{182}Ta, which has a half-life of \SI{114.61}{days}~\cite{be2011table}. Regarding radioactive waste management, Ta2.5W and C103 both produce \textsuperscript{3}H in similar amounts, while \textsuperscript{94}Nb is also produced in C103, and this poses significantly higher waste-disposal challenges due to its half-life of 20,300 years~\cite{adams1995national}. This could be a significant obstacle for the use of Nb materials as cladding for particle-producing targets.

\section{Conclusions}
This diffusion-bonding study of Nb, Nb1Zr, and C103 included the fabrication and testing of prototypes clad with these materials, and thermomechanical simulations were also conducted to validate their use as cladding materials for the BDF target. The results confirmed the ability of the Nb materials to form diffusion bonds with TZM and W, demonstrating good thermal contact at the bonding interfaces and high interface strengths. The use of a Ta-foil interlayer resulted in higher interface strengths for the TZM specimens and lower interface strengths for the W specimens. Specifically, C103 exhibited the highest interface strengths and the highest safety factors based on the results of the thermomechanical simulations. As such, C103 is considered to be superior to Ta2.5W as a cladding material for the BDF target due to its better mechanical properties, higher safety factors, and lower activation and decay heat. Nonetheless, the production of \textsuperscript{94}Nb could present a significant obstacle in terms of radiation waste management due to its extremely long half-life. In summary, the Nb materials considered in this study show promising results. As a next step, full-scale Nb-material-clad blocks should be produced and characterized.

\section*{Declaration of competing interest}
The authors declare that they have no known competing financial interests or personal relationships that could have appeared to influence the work reported in this paper.

\bibliographystyle{elsarticle-num}
\bibliography{Library}

\end{document}